\documentclass{sf2a-conf2013}
\usepackage{graphicx}
\usepackage{hyperref}
\usepackage[]{natbib}  
\usepackage{epstopdf}

\def\BibTeX{{\rm B\kern-.05em{\sc i\kern-.025em b}\kern-.08em
    T\kern-.1667em\lower.7ex\hbox{E}\kern-.125emX}}
\bibpunct{(}{)}{;}{a}{}{,}  

\begin{document}

\TitreGlobal{SF2A 2013}


\title{Perspectives for the study of gas in protoplanetary disks and accretion/ejection phenomena in young stars 
with the near-IR spectrograph SPIROU at the CFHT}

\runningtitle{The circumstellar environment of young stars traced by SPIROU.}

\author{A. Carmona}\address{UJF-Grenoble 1 / CNRS-INSU, 
  Institut de Plan\'etologie et d'Astrophysique de Grenoble (IPAG) UMR 5274, Grenoble, F-38041, France}

\author{J. Bouvier$^1$}

\author{X. Delfosse$^1$}




\setcounter{page}{237}


\maketitle


\begin{abstract}
Near-IR atomic and molecular transitions are powerful tools 
to trace the warm and hot gas in the circumstellar environment of young stars.
Ro-vibrational transitions of H$_2$ and H$_2$O, and overtone transitions of CO at 2 $\mu$m
centered at the stellar velocity trace hot (T$\sim$~1500 K) gas in the inner few AU of protoplanetary disks.
H$_2$ near-IR lines displaying a blueshift of a few km/s probe molecular disk winds.
H$_2$ lines presenting blueshifts of hundreds of km/s reveal hot shocked gas in jets.
Atomic lines such as the HeI line at 10830~\AA~ and the Hydrogen Paschen $\beta$ and Brakett $\gamma$ lines 
trace emission from accretion funnel flows and atomic disk winds.
Bright forbidden atomic lines in the near-IR of species such as [Fe~II], [N~I], [S~I], [S~II], and [C~I]  trace atomic and ionized
material in jets.  
The new near-IR high resolution spectrograph SPIROU planned for the Canada France Hawaii Telescope will 
offer the unique capability of combining 
high-spectral resolution (R$\sim$75000) with a large wavelength coverage (0.98 to 2.35 $\mu$m) in one single exposure.
This will provide us with the means of probing accretion funnel flows, winds, jets, and 
hot gas in the inner disk simultaneously. 
This opens the exiting possibility of investigating 
their combined behavior in time by the means of monitoring observations and systematic surveys.
SPIROU will be a powerful tool to progress our understanding of the connexion between the accretion/ejection process, disk evolution, and planet formation.  

\end{abstract}



\section{Introduction}
The extraordinary diversity of the extrasolar planetary systems discovered
has renewed our interest in understanding the planet formation process.
Planets form in the circumstellar disks that surround stars in their pre-main sequence. 
Such ``protoplanetary disks" are composed of gas (99\%) and dust, they extend typically hundreds of 
AU, have masses of a few percent of the central star, and have a lifetime on average of a few 
million years.
What is their geometry?
What are their density, temperature, and chemical structure?
What are the principal physical and chemical mechanisms setting that structure?
What are their dynamics? 
How do they dissipate?
To determine how disks are is crucial to understand planet formation.

Young stars are complex dynamical entities.
In addition to disks, they display jets, winds, and magnetospheric accretion funnel flows.
In fact the presence of collimated outflows and the presence of accretion disks appear to be inseparable
phenomena.
How are these different components interrelated?
How is the disk structure linked to accretion/ejection processes?
How do jets and accretion columns affect the radiation field interacting with the disk?
How do accretion/ejection phenomena affect the planet formation process?
How does this complex system involve in time?
What is their impact in the stellar evolution?
To understand the circumstellar environment of young stars
as a whole, 
it is a crucial aspect to understand early stellar evolution and the origin of planetary systems.
  
\section{Probing the gas in the circumstellar environment of young stars with SPIROU}

The dust content of circumstellar disks is traced by its thermal and scattered light emission.
The gas content of circumstellar disks, winds, outflows, collimated jets, and accretion flows is investigated
employing spectral lines. 
Atomic and molecular transitions trace regions at different temperatures and densities. 
Their  integrated fluxes, line shapes, velocity shifts, and line-ratios 
have the imprint the excitation mechanisms and the dynamics of the medium where the lines originate.
Different regions are traced at different wavelengths as a function of the temperature
of the medium producing the lines.
Cold gas ($T<$100 K) situated in the outer region of a protoplanetary disk (R$>$ 10 AU), 
or cold gas excited by shocks by outflows or jets are traced with
molecular transitions in the sub-mm and mm wavelengths ($\lambda>400~\mu$m).
Warm gas ($100<T<1000$ K) located in the surface layer of a disk between a few AU and 10 AU (i.e. the giant planet forming region)
will be probed by mid-IR ($8-100~\mu$m) and far-IR ($100-400~\mu$m) transitions. 
Hot gas (T$>$ 1000 K) in the inner few AU of disks (i.e. the terrestrial planet forming region), accretion funnel flows,
disk winds, and jets are traced by near-IR ($1-8~\mu$m) emission.

\begin{figure}[t!]
 \centering
 \includegraphics[width=1.0\textwidth,clip]{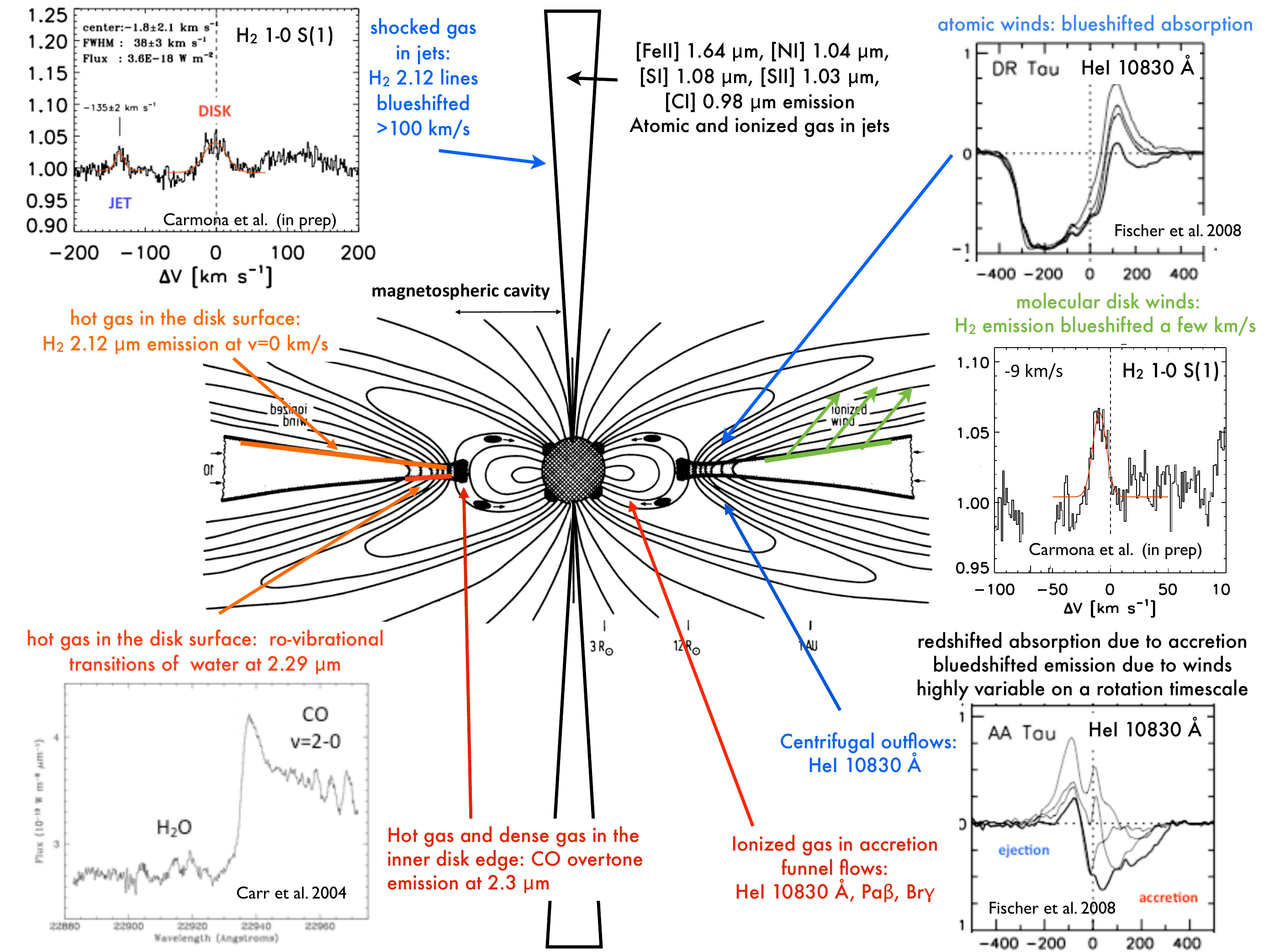}      
  \caption{Summary of diagnostics of gas in the circumstellar environment of young stars that will be covered by SPIROU.
  T Tauri star cartoon adapted from \citet{Camenzind1990}. }
  \label{Carmona:fig1}
\end{figure}

The keys to discern between the different origins of a line are its shape and velocity shift with respect to the central star (see Fig. \ref{Carmona:fig1}). 
Emission produced  by a circumstellar disk will {\it centered at the velocity of the star} and will produce the characteristic double peak 
profiles due to Keplerian rotation. 
The double peaked profile provide us information of the region in the disk emitting the line.
The lines are broader if the line is emitted closer to the star.
The line high velocity wings indicate the inner-most radius emitting the line.
The double peak separation provides us information about the outer radius.
The double peak separation and the line width tell us about the disk inclination.
Emission produced by a collimated jet, or a fast moving outflow, will be {\it blue-shifted hundreds of km/s}.
Emission produced by a disk wind will be {\it asymmetric and/or blue-shifted a few km/s}.
The presence of outflows and winds can also be seen as blueshifted absorption lines on the top emission component (i.e. P Cygni profile).
Accretion flows are traced by inverse P Cygni profiles (red-shifted absorption on the top of an emission component).
Emission from fast moving jets, winds, or outflows from the inner most disk can be traced already at low spectral resolution (R$\sim$2000 or 150 km/s),
however, to exploit the dynamical information encoded in the lines from of disks or disk winds, high-spectral resolution (R$>$10000) is a must.
Furthermore, in the case of disk emission, high-spectral resolution is required to separate the gas
lines from telluric absorption and detect weak emission lines on the top of the strong dust continuum.

SPIROU will trace in one exposure the $0.98-2.35$ $\mu$m region  at a spectral resolution 75000.
In this region, we can study 
H$_2$ near-IR emission, for example the 1-0 S (1), 1-0 S(0), and 2-1 S(1) ro-vibrational lines at 2.12, 2.22, and 2.24~$\mu$m \cite[e.g.][]{Bary2003,RamsayHowat2007,Carmona2011},
ro-vibrational water emission at  2.29 $\mu$m \cite[e.g.][]{Carr2004,Thi2005a}; and
CO overtone $\Delta v=2$ bandhead emission at 2.3~$\mu$m \cite[e.g.][]{Chandler1995,Najita1996,Thi2005b}.
These molecular near-IR transitions trace hot gas (T$>$1000 K) in the inner few AU of disks.
They probe different excitation mechanisms, temperatures, and densities. 
For example, near-IR H$_2$ lines are excited in the upper layers of the disk either by UV-radiation or X-rays and are sensitive to earth-masses of gas at $\sim$1000 K.
CO overtone emission, in contrast, requires very dense ($n_{\rm H}>10^{10} $ cm$^{-3}$) and hot ($T =2000-4000$ K) gas to be produced.
Furthermore, H$_2$ near-IR lines are  good tracers of hot shocked gas in outflows and collimated jets extending at hundreds of AU.
They have also the potential of tracing molecular disk winds excited by energetic radiation from the central source.
The near-IR window covered by SPIROU permits us to measure
atomic lines such as the HeI line at 10830~\AA~ and the Hydrogen Paschen $\beta$ (1.28 $\mu$m) and Brakett $\gamma$ (2.17 $\mu$m) lines.
These lines are primordial for the study of the magnetospheric cavity, the accretion/ejection process, and the interface between the disk 
and the star. These lines are powerful tools to study accretion funnel flows and atomic disk winds \cite[e.g.][]{Fischer2008}.
Bright forbidden atomic lines in the near-IR such as the [Fe II] lines at 1.64, 1.59,  1.53, 1.32, and 1.25 $\mu$m, 
the [N~I] lines at 1.04 $\mu$m, 
the [S~I] lines at 1.08 and 1.13 $\mu$m, 
the [S~II] lines at 1.03 $\mu$m,
and the  [C~I] lines at 0.98 $\mu$m
trace neutral and ionized material in jets \cite[e.g.][]{Giannini2006}.
In Figure~\ref{Carmona:fig1}, we present a cartoon summarizing the different diagnostics of the circumstellar environment of young stars covered by SPIROU. 

Current high-resolution (R$>$ 10000) near-IR spectrographs have limited wavelength coverage.
For example, 
the near-IR spectrograph CRIRES at ESO/VLT (R$\sim$90000)
covers in one exposure 0.02 $\mu$m. 
NIRSPEC at Keck (R$\sim$25000) covers 0.18~$\mu$m in one setting.
The unique capability of SPIROU would be to offer a coverage of 1.37 $\mu$m (0.98 to 2.35 $\mu$m) in one single setting with a resolution of 4 km/s.
This will provide us with the means of measuring diagnostics of accretion, winds, jets, and disks, all simultaneously in one shot. 
SPIROU is expected to reach a S/N $>$100 in a 2 km/s pixel in 1h of exposure for objects with magnitudes up to J=12 and K=11,
thus a 5 $\sigma$ flux sensitivity of $\sim$ 10$^{-19} $ W m$^{-2}$ in a line of width  8 km/s .
The sensitivity and large spectral coverage provided by SPIROU
will permit us to investigate in detail a large number of sources employing short exposure times. 
This opens the exiting possibility of investigating the evolution in time of disks, winds, and jets simultaneously
by the means of monitoring observations.
SPIROU will be a powerful tool to progress our understanding of the connexion between the accretion/ejection process, disk evolution, and planet formation.  

\begin{acknowledgements}
{\it Acknowledgements:} A. Carmona acknowledges funding from the
Agence Nationale pour la Recherche (ANR) of France under contract
ANR-2010-JCJC-0504-01.
\end{acknowledgements}

\bibliographystyle{aa}  
\bibliography{Carmona} 

\end{document}